\documentclass[letterpaper]{article} 
\usepackage{aaai2027}  
\usepackage[hyphens]{url}  
\usepackage{graphicx} 
\usepackage{natbib}  
\usepackage{caption} 
\usepackage{newfloat}
\usepackage{listings}
\usepackage{float}
\DeclareCaptionStyle{ruled}{labelfont=normalfont,labelsep=colon,strut=off} 
\floatstyle{ruled}
\newfloat{listing}{tb}{lst}{}
\floatname{listing}{Listing}

\usepackage{booktabs}

\usepackage{algorithm}
\usepackage{booktabs}
\usepackage{multirow}
\usepackage{adjustbox}
\usepackage{makecell}
\usepackage{amsmath}
\usepackage[most]{tcolorbox}
\usepackage{xcolor}

\usepackage{newfloat}
\usepackage{listings}
\DeclareCaptionStyle{ruled}{labelfont=normalfont,labelsep=colon,strut=off} 
\floatstyle{ruled}
\newfloat{listing}{tb}{lst}{}
\floatname{listing}{Listing}
\usepackage[most]{tcolorbox}
\usepackage{xcolor}
\usepackage{listings}
\usepackage{algorithm}
\usepackage{algpseudocode}
\usepackage{booktabs}
\usepackage[most]{tcolorbox}
\usepackage{xcolor}
\usepackage{listings}
\usepackage{fontawesome5} 
\usepackage{xspace}
\definecolor{SpecBlue}{HTML}{2F5DA8}
\definecolor{SpecBg}{HTML}{F7F9FC}
\definecolor{SpecCodeBg}{HTML}{FBFCFE}
\definecolor{SpecBorder}{HTML}{C9D6EA}
\definecolor{SpecComment}{HTML}{6A737D}
\definecolor{SpecKeyword}{HTML}{005CC5}
\definecolor{SpecString}{HTML}{032F62}
\usepackage{xcolor}
\usepackage[most]{tcolorbox}
\usepackage{xcolor}
\usepackage{listings}
\usepackage{graphicx}
\usepackage{subcaption}
\definecolor{SpecNavy}{HTML}{1F3A5F}
\definecolor{SpecTitleBg}{HTML}{DDE8F7}
\definecolor{SpecCodeBg}{HTML}{F3F6FB}
\definecolor{SpecBorder}{HTML}{6F8FB8}
\definecolor{SpecComment}{HTML}{4F5B66}
\definecolor{SpecKeyword}{HTML}{1D4E89}
\definecolor{SpecString}{HTML}{8A4B08}
\definecolor{SpecLine}{HTML}{8B98A8}

\newcommand{\score}[2]{%
#1\textcolor{gray}{\scriptsize$\pm$#2}%
}

\newcommand{\bestscore}[2]{%
\textbf{#1}\textcolor{gray}{\scriptsize$\pm$#2}%
}
\lstdefinestyle{archspec}{
    language=Python,
    basicstyle=\ttfamily\scriptsize,
    keywordstyle=\color{SpecKeyword}\bfseries,
    commentstyle=\color{SpecComment}\itshape,
    stringstyle=\color{SpecString},
    showstringspaces=false,
    breaklines=true,
    columns=fullflexible,
    keepspaces=true,
    frame=none,
    numbers=left,
    numberstyle=\tiny\color{SpecLine},
    numbersep=6pt,
    xleftmargin=1.3em,
    aboveskip=0pt,
    belowskip=0pt
}

\newtcolorbox{SummaryBox}{
  enhanced,
  breakable,
  colback=blue!5,        
  colframe=blue!40,      
  boxrule=0pt,           
  leftrule=3pt,          
  rightrule=0pt,
  toprule=0pt,
  bottomrule=0pt,
  arc=3pt,               
  left=1pt,
  right=1pt,
  top=1pt,
  bottom=1pt,
}

\def\ourmethod{\textsc{CodeSpec}\xspace}

\title{\ourmethod: Dual Executable Specifications for Agentic Long-Horizon Feature Development}

\author{
    Peiding Wang\textsuperscript{\rm 1},
    Li Zhang\textsuperscript{\rm 1},
    Fang Liu\textsuperscript{\rm 1}\corresponding,
    Taichuan Li\textsuperscript{\rm 1},
    Yinghao Zhu\textsuperscript{\rm 2} \\
}
\affiliations{
    \textsuperscript{\rm 1}State Key Laboratory of Complex \& Critical Software Environment, \\
School of Computer Science and Engineering, Beihang University, China \\
    \textsuperscript{\rm 2}School of Computing and Data Science, The University of Hong Kong\\
    {\tt \{wangpeiding, fangliu\}@buaa.edu.cn}

}

\begin{document}

\maketitle

\begin{abstract}
LLM-based code agents have advanced repository-level software development through iterative interaction with codebases and tools. However, feature development requires integrating new behaviors into existing architectures through coherent cross-component functional chains. Existing agents typically derive such chains through free-form reasoning, often producing unreliable feature designs with incomplete functional chains. Moreover, textual designs are difficult to verify and enforce, making it challenging to maintain design–implementation consistency throughout long-horizon development.
We propose \ourmethod{}, a dual executable specification method for repository-level feature development. It builds reliable functional chains from evidence pairing sub-requirement semantics with repository architectures, then compiles them into complementary architecture and behavior specifications that check chain completeness and correctness while preserving design–implementation consistency over long interactions. On FeatureBench, which targets feature development in existing repositories, \ourmethod{} achieves 70.7\%, 55.0\%, and 49.9\% pass rates under DeepSeek-V4-Pro, outperforming representative baselines such as Claude Code. Results on the repository generation benchmark NL2Repo-Bench further demonstrate its generalizability. 
The code and data are available at \url{https://github.com/zhu-zhu-ding/CodeSpec}.
\end{abstract}

\section{Introduction}

With the rapid advancement of large language models, automated software development has evolved from standalone function generation~\cite{humaneval,mbpp} to repository-level tasks~\cite{deveval,swebench}. Feature development is particularly important because real-world software systems are continuously extended to satisfy evolving requirements~\cite{swedev,feabench,featurebench}. Unlike standalone generation, implementing a new feature requires integrating new behaviors into an existing architecture. Agents must determine not only which code entities should be created or modified, but also how these entities should interact across components. Feature development therefore commonly requires the design stage that organizes these entities and interactions into a coherent functional chain from feature entry points to observable behaviors.

\begin{figure}[t]
    \centering
    \includegraphics[width=\linewidth]{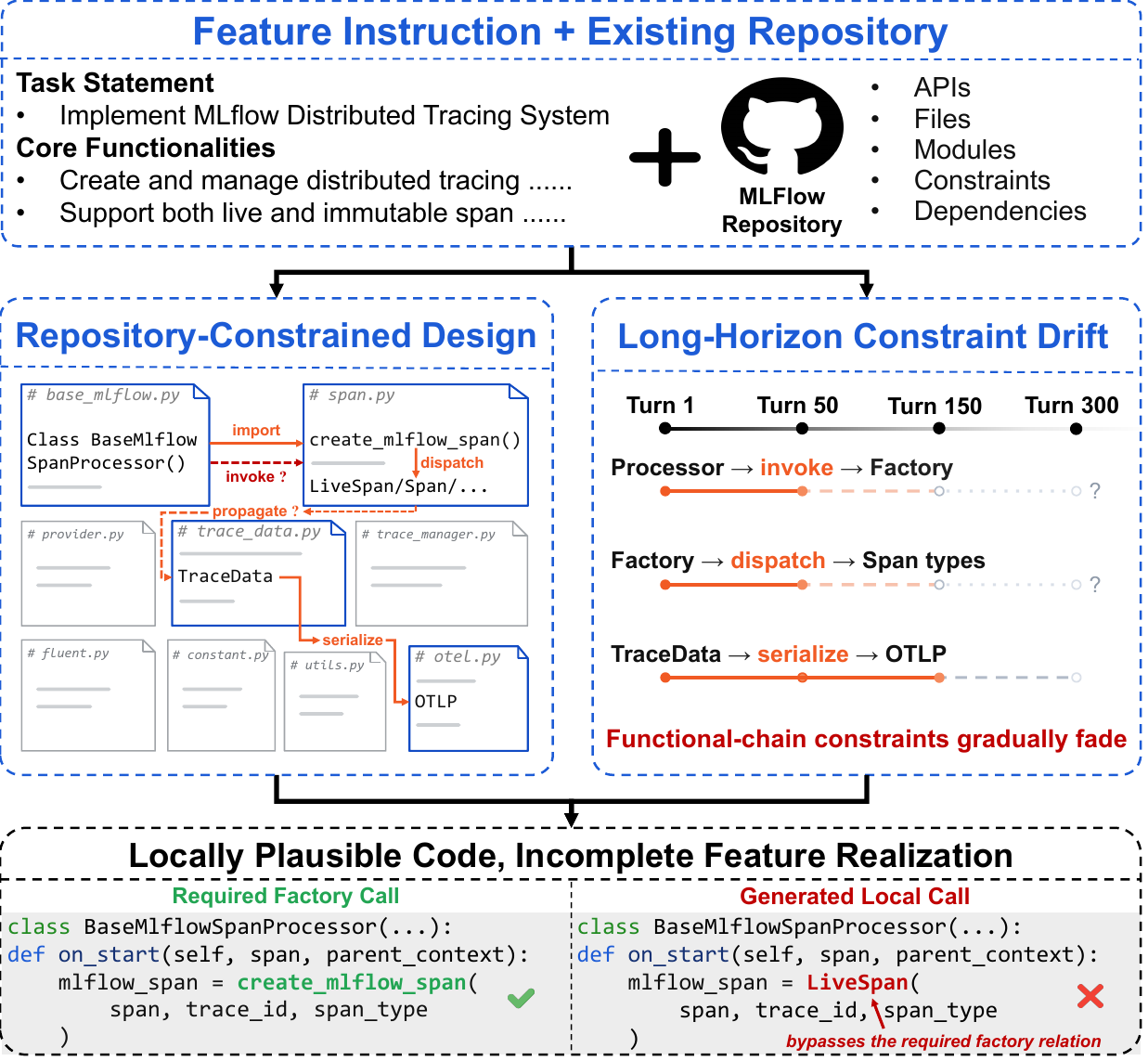}
    \caption{Feature development requires agents to derive repository-consistent functional chains and preserve them throughout implementation. Missing a critical architectural relation, such as invoking factory method \textit{create\_mlflow\_span()}, may break the end-to-end feature behavior.}
    \label{fig:motivation}
    \vspace{-0.8cm}
\end{figure}

The repository-level nature of feature development makes LLM-based coding agents a promising paradigm. Representative agents, including Claude Code~\cite{claudecode}, Codex~\cite{codex}, SWE-agent~\cite{sweagent}, and OpenHands~\cite{openhands}, rely on the agent’s reasoning capabilities to implicitly plan development through natural-language interactions while iteratively generating and repairing code. Other approaches introduce explicit design processes: MetaGPT~\cite{metagpt} and ChatDev~\cite{chatdev} produce natural-language artifacts describing module responsibilities and collaboration workflows, while RTADev~\cite{liurtadev} performs architecture planning before implementation. However, both free-form implicit planning and explicit textual design provide limited support for deriving reliable repository-grounded functional chains and enforcing them throughout long-horizon development.

As illustrated in Figure~\ref{fig:motivation}, existing agents face two closely related challenges:

\begin{itemize}
\item \textbf{Unreliable functional-chain design.}
Existing agents typically design features through free-form reasoning, which can deviate from repository architecture constraints and overlook the call and dependency relations among code entities. Consequently, they may derive functional chains that are incomplete or inconsistent with the repository.

\item \textbf{Weak long-horizon enforcement.}
Functional designs are typically represented as free-form textual plans, which are difficult to verify and enforce during iterative exploration, editing, and repair. As a result, the implementation may gradually diverge from the intended functional chain, leading to inconsistencies between design and realization.

\end{itemize}

To address these challenges, we propose \ourmethod{}, a dual executable specification method for agentic long-horizon feature development. It constructs repository-grounded functional-chains for feature sub-requirements and compiles each chain into complementary architecture and behavior executable specifications. \ourmethod{} consists of two key components:

\noindent\textbf{1) Evidence-Grounded Functional Chain} derives traceable feature realization paths by identifying relevant functional units and connecting them with paired requirement semantics and repository as evidence, such as design patterns, call relations, and dependencies. This reduces free-form architectural reasoning and grounds each transition in the existing repository to improve the design reliability.

\noindent\textbf{2) Dual Executable Specification Compilation} converts each functional chain into two executable forms: an architecture specification that represents its architecture realization and a behavior specification that represents its expected execution. They complementarily verify whether the chain is complete and whether it behaves correctly, thereby preserving design--implementation consistency throughout long-horizon development.

Experiments on FeatureBench~\cite{featurebench} show that \ourmethod{} achieves 70.7\%, 55.0\%, and 49.9\% \%Passed on the Lite, Fast, and Full splits, consistently outperforming coding-agent and architecture-planning baselines under the same DeepSeek-V4-Pro backbone. Results with GPT-5.4-mini and NL2Repo-Bench~\cite{nl2repobench} which evaluates greenfield repository development from natural-language requirements further demonstrate robustness across models and generalization to greenfield repository generation. Ablation and further analyses confirm the effectiveness of evidence-grounded functional chains and the complementary executable specifications, particularly on complex and long-horizon tasks, with moderate additional cost.

Our main contributions are summarized as follows:
\begin{itemize}
    \item We propose evidence-grounded functional-chain reasoning that pairs task semantics with repository architecture evidence at each transition, enabling agents to derive reliable cross-component realization paths from feature entry points to observable behaviors.

    \item We introduce dual executable specifications that represent the architecture realization and expected execution of each functional chain, jointly verifying whether the chain is complete and behaves correctly while maintaining design--implementation consistency throughout long-horizon development.

    \item We conduct extensive experiments on FeatureBench and NL2Repo with different backbone models, demonstrating the effectiveness, robustness, generalizability, and favorable quality--cost trade-off of \ourmethod{} for long-horizon feature development.
\end{itemize}

\begin{figure*}[h]
    \centering
    \setlength{\abovecaptionskip}{0.1cm}
    \includegraphics[width=\linewidth]{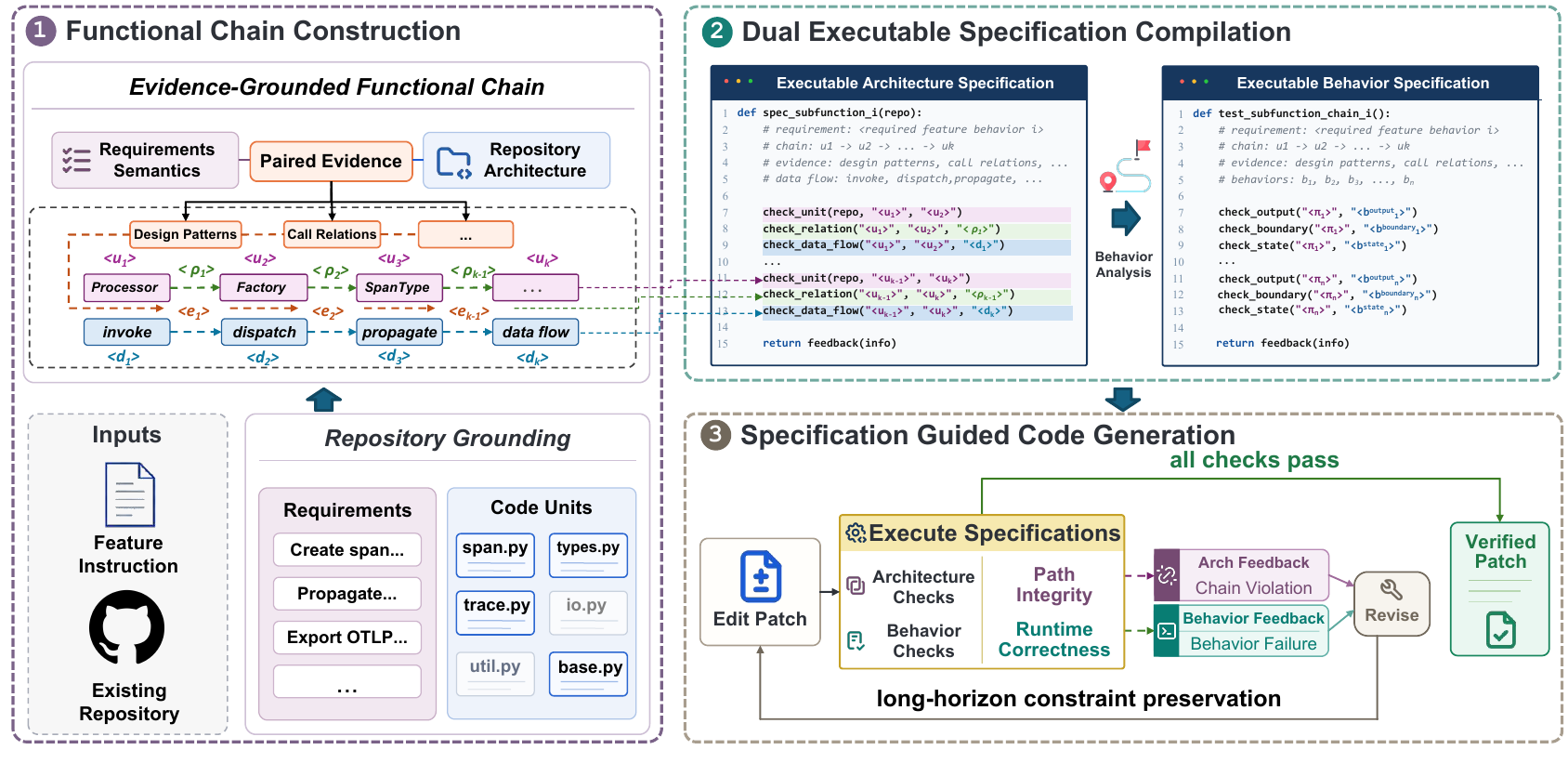}
    \caption{The overview of \ourmethod.}
	\label{fig:main_pipline}

\end{figure*}
\section{Related Work}

\subsection{Repository-level Feature Development Benchmarks}

Code generation research has evolved from standalone function synthesis~\cite{humaneval, mbpp} to repository-level software development, where models need to understand project context and generate patches in real codebases. Benchmarks such as DevEval~\cite{deveval} and SWE-bench~\cite{swebench} evaluate such repository-level capabilities through realistic software engineering tasks. More recent feature-level benchmarks, including FEA-Bench~\cite{feabench}, and FeatureBench~\cite{featurebench}, further focus on practical feature development, where a high-level instruction may require coordinated changes and generation across multiple functions, files, classes, and interfaces. These tasks usually involve a full development process of design, implementation, testing, execution, and repair, making one-shot generation insufficient and motivating coding agents that interact with repositories and development environments over multiple steps.

\subsection{LLM-based Coding Agents}
LLM-based coding agents have become a popular paradigm for repository-level software development~\cite{codemem,efficientedit,codeifbench}. Unlike one-shot generation methods, they can autonomously invoke tools, inspect files, text plan, edit programs, and iteratively repair failures. Representative coding agents and assistants include Claude Code~\cite{claudecode}, Codex~\cite{codex}, SWE-agent~\cite{sweagent}, and OpenHands~\cite{openhands}. In addition, development-oriented agent frameworks such as MetaGPT~\cite{metagpt}, ChatDev~\cite{chatdev} and RTADev~\cite{liurtadev}  organize software development into structured stages and use textual architecture designs to coordinate software development. Despite their progress, existing agents still mainly maintain design information as free-form textual plans, which makes it difficult to explicitly preserve functional chains and architecture constraints during long-horizon feature development. 

\section{Method}
\subsection{Task Definition}

In this section, we define the \textbf{feature development} task. Given a natural-language feature instruction $I$ and an existing repository ${R}$, an LLM-based coding agent generates a patch $P$ to get the updated repository ${R'}$ which preserving existing functionality and satisfying the instruction $I$:
\begin{equation}
    P = \operatorname{LLM}(I, {R}), 
    \qquad
    {R}' = {R} \oplus P .
\end{equation}
\subsection{Overview}
As shown in Figure~\ref{fig:main_pipline}, \ourmethod constructs reliable functional chains using evidence that pairs sub-requirement semantics with the corresponding repository architecture. It then compiles each chain into dual executable specifications: an architecture specification representing its structural realization and a behavior specification representing its expected execution. These specifications complementarily verify whether the chain is complete and behaves correctly, thereby maintaining design--implementation consistency throughout long-horizon feature development.






\subsection{Evidence-Grounded Functional Chain}

A feature is realized through coordinated interactions among multiple repository components rather than isolated code modifications. However, feature instructions usually describe desired behaviors without specifying which components should participate or how they should interact. To bridge this gap, \ourmethod{} constructs an evidence-grounded functional chain for each sub-function by pairing requirement semantics with repository architecture evidence at every reasoning step.

Given a feature instruction $I$, the agent first decomposes it into a set of sub-requirements:

\begin{equation}
    \mathrm{LLM}(I)
    =
    \{r_1,r_2,\ldots,r_m\}.
\end{equation}

Each requirement $r_i$ represents a specific behavior that the feature should achieve. For each requirement, the agent identifies candidate implementation units that need to be modified or introduced:

\begin{equation}
    U_{r_i}=\{u_1,u_2,\ldots,u_n\}.
\end{equation}

These units provide candidate building blocks, but selecting relevant units alone does not yield a valid feature design; the key is determining how they should be connected to realize each obligation. \ourmethod{} therefore prompts the LLM to perform evidence-driven chain reasoning by pairing requirement evidence, which explains why a transition is needed from requirements, with repository evidence, which shows how it is supported by existing design patterns, call relations, or repository architectures. The resulting functional chain is an ordered sequence in which each adjacent pair of units $u_i^k$ and $u_i^{k+1}$ is linked according to its supporting evidence $e_i^k$:
\begin{equation}
    c_i=
    \{
    u_i^1,e_i^1,
    u_i^2,
    \ldots,
    u_i^{k},e_i^{k},u_i^{k+1}
    \}.
\end{equation}
Here, $u_i^k$ denotes a functional unit involved in realizing requirements $r_i$, and $e_i^k$ denotes the evidence supporting the transition from $u_i^k$ to $u_i^{k+1}$. Each evidence $e_i^k$ pairs requirement semantics with repository evidence, ensuring that adjacent units are connected through traceable, repository-supported reasoning rather than free code relations.

\subsection{Dual Executable Specification Compilation}

The constructed functional chains explicitly describe how individual feature requirement should be realized across repository components. However, retaining these chains as textual design artifacts is unsuitable for long-horizon development, because agents may gradually overlook pre-designed key information during long-term interactions. Therefore, we compile each functional chain into two complementary executable specifications: the architecture specification that verifies path integrity and the behavior specification that verifies runtime correctness along the path.

\subsubsection{Executable Architecture Specification}

Given a sub-requirement $r_i$ and its functional chain $c_i$, \ourmethod{} instantiates an executable architecture specification $s_i^{\mathrm{arch}}$ using the template $\tau_{\mathrm{arch}}$ shown in Figure~\ref{fig:main_pipline}:

\begin{equation}
    s_i^{\mathrm{arch}}
    =
    \mathrm{LLM}_{\mathrm{arch}}
    (R,r_i,c_i;\tau_{\mathrm{arch}}).
\end{equation}

Inspired by architecture conformance checking and program analysis
~\cite{arch1,arch2,arch3},
we operationalize path integrity using three representative dimensions:
functional units, architectural relations, and data flows. The template maps the units, relations, and data flows in $c_i$ into three types of checks:

\begin{equation}
\begin{aligned}
s_i^{\mathrm{arch}} ={}&
    \left\{
    \mathrm{CheckUnit}(u_i^j)
    \right\}_{j=1}^{k_i}
    \\
    &\cup
    \left\{
    \mathrm{CheckRelation}
    (u_i^j,u_i^{j+1},\rho_i^j)
    \right\}_{j=1}^{k_i-1}
    \\
    &\cup
    \left\{
    \mathrm{CheckDataFlow}
    (u_i^j,u_i^{j+1},d_i^j)
    \right\}_{j=1}^{k_i-1}.
\end{aligned}
\end{equation}

Here, $\rho_i^j$ and $d_i^j$ are jointly inferred from the requirement and repository evidence in $e_i^j$, representing the intended architectural relation and data-flow operation between $u_i^j$ and $u_i^{j+1}$, respectively. \textsc{CheckUnit} verifies that each required unit exists, \textsc{CheckRelation} checks whether adjacent units preserve the intended relation $\rho_i^j$, and \textsc{CheckDataFlow} verifies the required data state $d_i^j$ between them. 

During implementation, the specification is executed against the updated repository. Missing units, broken relations, or interrupted data flows are reported as localized architecture violations. In this way, executable architecture specifications preserve the path integrity of each functional chain.

\subsubsection{Executable Behavior Specification}

Given the sub-requirement $r_i$, its functional chain $c_i$, and the corresponding architecture specification $s_i^{\mathrm{arch}}$, \ourmethod{} jointly analyzes the intended path structure and architectural constraints to identify behavior-observation subpaths, and instantiates an executable behavior specification using the template $\tau_{\mathrm{beh}}$ shown in Figure~\ref{fig:main_pipline}:

\begin{equation}
    s_i^{\mathrm{beh}}
    =
    \mathrm{LLM}_{\mathrm{beh}}
    (r_i,c_i,s_i^{\mathrm{arch}};\tau_{\mathrm{beh}}).
\end{equation}

Inspired by established property-based, category-partition, and model-based testing techniques~\cite{beh1,beh2,beh3}, we operationalize behavioral obligations using three representative dimensions: observable outputs, boundary conditions, and state transitions. The template derives three types of behavioral checks at multiple observation scopes along the functional chain:

\begin{equation}
\begin{aligned}
s_i^{\mathrm{beh}}=\{&
    \mathrm{CheckOutput}
    (\pi_i^q,b_{i,q}^{\mathrm{output}}),\\
    &\mathrm{CheckBoundary}
    (\pi_i^q,b_{i,q}^{\mathrm{boundary}}),\\
    &\mathrm{CheckState}
    (\pi_i^q,b_{i,q}^{\mathrm{state}})
\}_{q=1}^{n_i}.
\end{aligned}
\end{equation}

Here, $n_i$ denotes the number of behavioral observation scopes derived for $c_i$, and $\pi_i^q \subseteq c_i$ denotes the $q$-th scope, which may cover a functional unit, a transition, or a subpath. The checks examine observable outcomes, including returned values and external effects, boundary and exceptional conditions, and state transitions at multiple positions along the functional chain. Thus, the behavior specification examines multiple intermediate and final behaviors rather than only the final output.

During implementation, violated expectations are reported as localized behavior violations. In this way, executable behavior specifications complement architecture specifications by preserving runtime correctness along each functional chain.

\subsection{Specification Guided Code Generation}

For all sub-requirements, the generated architecture and behavior specifications are aggregated as:

\begin{equation}
    S^{\mathrm{arch}}
    =
    \bigcup_{i=1}^{m}s_i^{\mathrm{arch}},
    \qquad
    S^{\mathrm{beh}}
    =
    \bigcup_{i=1}^{m}s_i^{\mathrm{beh}}.
\end{equation}

The agent implements the feature under the joint constraints of $S^{\mathrm{arch}}$ and $S^{\mathrm{beh}}$. Architecture feedback identifies missing units, broken relations, or interrupted data flows, while behavior feedback reports incorrect runtime behaviors along the functional chains. The agent incorporates both signals into subsequent edits until all specifications pass or the interaction budget $K$ is exhausted.

\begin{algorithm}[h]
\caption{Specification Guided Code Generation}
\label{alg:code_generator}
\begin{algorithmic}[1]
\Statex \textbf{Input:} Instruction $I$, repository $R$, specifications $S^{\mathrm{arch}}$ and $S^{\mathrm{beh}}$, budget $K$
\Statex \textbf{Output:} Patch $P$

\State $P,F_{\mathrm{arch}},F_{\mathrm{beh}} \gets \emptyset$
\State $k \gets 0$

\While{$k<K$}
    \State $\Delta P \gets
    \mathrm{LLM}_{\mathrm{code}}
    (I,R,S^{\mathrm{arch}},S^{\mathrm{beh}},
    P,F_{\mathrm{arch}},F_{\mathrm{beh}})$
    \State $P \gets P \oplus \Delta P$
    \State $R' \gets R \oplus P$
    \State $F_{\mathrm{arch}}
    \gets \mathrm{ExecArch}(R',S^{\mathrm{arch}})$
    \State $F_{\mathrm{beh}}
    \gets \mathrm{ExecBeh}(R',S^{\mathrm{beh}})$

    \If{$\mathrm{Pass}(F_{\mathrm{arch}})
    \land \mathrm{Pass}(F_{\mathrm{beh}})$}
        \State \textbf{break}
    \EndIf

    \State $k \gets \mathrm{InteractCount}()$
\EndWhile

\State \textbf{return} $P$
\end{algorithmic}
\end{algorithm}

Here, $\mathrm{ExecArch}$ verifies the integrity of the functional chains, whereas $\mathrm{ExecBeh}$ verifies their runtime behaviors. Their complementary feedback jointly preserves path integrity and behavioral correctness throughout long-horizon implementation.

\section{Experimental Setups}

\begin{table*}[t]
\centering
\small
\setlength{\tabcolsep}{4.5pt}
\renewcommand{\arraystretch}{1.15}
\begin{adjustbox}{max width=\textwidth}
\begin{tabular}{lccccccccc}
\toprule
\multirow{2}{*}{\textbf{Agent Scaffold}}
& \multicolumn{3}{c}{\textbf{Lite}}
& \multicolumn{3}{c}{\textbf{Fast}}
& \multicolumn{3}{c}{\textbf{Full}} \\
\cmidrule(lr){2-4}
\cmidrule(lr){5-7}
\cmidrule(lr){8-10}
& \textbf{\%Passed}
& \textbf{\#Resolved}
& \textbf{Avg. Cost (\$)}
& \textbf{\%Passed}
& \textbf{\#Resolved}
& \textbf{Avg. Cost (\$)}
& \textbf{\%Passed}
& \textbf{\#Resolved}
& \textbf{Avg. Cost (\$)} \\
\midrule
OpenHands
& \score{60.0}{5.7} & 6  & 0.15
& \score{41.4}{3.6} & 10  & 0.13
& \score{37.1}{2.7} & 18  & 0.11 \\

Claude Code
& \score{59.8}{5.4} & 6 & 0.11
& \score{47.0}{3.8} & 14 & 0.12
& \score{41.1}{2.8} & 23 & 0.10 \\

Mini-SWE-Agent
& \score{62.6}{6.0} & 7 & 0.09
& \score{49.6}{4.0} & 24 & 0.09
& \score{43.4}{2.8} & 25 & 0.09 \\

RTADev
& \score{65.3}{5.6} & 6  & 0.17
& \score{49.7}{3.8} & 15 & 0.16
& \score{46.1}{2.8} & 26 & 0.15 \\

\ourmethod
& \bestscore{70.7}{5.1} & \textbf{9}  & 0.18
& \bestscore{55.0}{3.8} & \textbf{17} & 0.19
& \bestscore{49.9}{2.8} & \textbf{28} & 0.16 \\
\bottomrule
\end{tabular}
\end{adjustbox}
\caption{Comparison under the DeepSeek-V4-Pro (max thinking) backbone model on FeatureBench. The Lite, Fast, and Full splits contain 30, 100, and 200 tasks, respectively. 
Bold numbers indicate the best performance.}
\label{tab:featurebench_same_model}
\end{table*}








\begin{table}[t]
\centering
\small
\setlength{\tabcolsep}{6pt}
\renewcommand{\arraystretch}{1.15}


\begin{tabular}{lccc}
\toprule
\textbf{Agent Scaffold}
& \textbf{\%Passed}
& \textbf{\#Resolved}
& \textbf{Avg. Cost (\$)}
\\
\midrule

Mini-SWE-Agent
& \score{47.5}{6.0}
& 2
& 0.7
\\

OpenHands
& \score{48.1}{5.9}
& 3
& 1.2
\\

RTADev
& \score{45.9}{5.7}
& 3
& 1.3
\\

Codex
& \score{51.7}{6.2}
& 4
& 0.7 \\

\ourmethod
& \bestscore{54.7}{5.8}
& \textbf{5}
& 1.3\\

\bottomrule
\end{tabular}
\caption{Comparison of different coding agents with the same GPT-5.4-mini backbone (medium thinking) on the FeatureBench Lite split.}
\label{tab:gpt54mini_agent}
\end{table}

\subsection{Research Questions}

We evaluate \ourmethod{} through the following research questions:

\begin{itemize}
    \item \textbf{RQ1: Overall Performance.}
    How does \ourmethod{} compare with existing coding agents on repository-level feature development tasks?

    \item \textbf{RQ2: Ablation Study.}
    How does each component contribute to the \ourmethod{}’s performance?

    \item \textbf{RQ3: Executable vs. Textual Specifications.}
    Do executable specifications preserve design constraints more effectively than textual specifications?

    \item \textbf{RQ4: Greenfield Repository Development Generalization.}
    Can \ourmethod{} generalize from repository evolution to greenfield repository generation?
\end{itemize}

\subsection{Benchmarks and Metrics}

We evaluate \ourmethod{} on FeatureBench~\cite{featurebench}, a repository-level feature development benchmark requiring coordinated modifications across mutli-interfaces. Each problem averages 4.8k words and requires about 800 lines of valid  code, compared with 0.2k words and about 32 lines in SWE-Bench\cite{swebench}. It contains three splits: \textit{Lite (30)}, \textit{Fast (100)}, and \textit{Full (200)}.

Following FeatureBench, we report \textbf{\%Passed}, the average proportion of fail-to-pass tests passed across tasks, and \textbf{Resolved \#}, the number of tasks for which all required tests pass. The former measures partial implementation progress, while the latter measures complete task resolution. We also reporte the average cost metric for each task: \textbf{Avg.Cost (\$)}.

\subsection{Baselines and Implementation Details}

We compare \ourmethod{} with five representative coding-agent scaffolds:

\begin{itemize}
    \item \textbf{OpenHands}~\cite{openhands}, a general-purpose agent supporting repository exploration, code editing, execution, and iterative repair.

    \item \textbf{Mini-SWE-Agent}~\cite{sweagent}, a lightweight agent based on a minimal repository interaction and repair loop.

    \item \textbf{Claude Code}~\cite{claudecode}, a widely adopted industrial coding agent for autonomous repository editing, execution, and debugging.

    \item \textbf{Codex}~\cite{codex}, an OpenAI coding agent supporting repository-level reasoning, tool use, code modification, and iterative refinement.

    \item \textbf{RTADev}~\cite{liurtadev}, an architecture-first agent that uses textual architecture designs to guide development;
\end{itemize}

For controlled comparison, all methods (except for Codex) use DeepSeek-V4-Pro as the backbone model. We additionally evaluate GPT-5.4-mini to examine model-level generalization. All methods share the same repository environments, evaluation scripts, and interaction budgets. We bootstrap task-level results 10,000 times and report the resulting mean and standard deviation.

\section{Experimental Results}
\subsection{RQ1: Overall Performance}

As shown in Table~\ref{tab:featurebench_same_model}, \ourmethod consistently achieves the best performance across all FeatureBench splits under the same DeepSeek-V4-Pro backbone. On the Lite split, \ourmethod reaches 70.7\% \%Passed and resolves 9 tasks, outperforming the strongest baseline RTADev by 5.4 percentage points in \%Passed and 3 resolved tasks. Similar improvements are observed on the Fast and Full splits, where \ourmethod improves \%Passed over RTADev by 5.3 and 3.8 percentage points, respectively, while achieving the highest number of resolved tasks. Compared with general-purpose coding agents such as OpenHands and Mini-SWE-Agent, \ourmethod obtains larger gains across all splits. Unlike textual plans, which remain passive context and can be overlooked during extended interactions, executable specifications continuously expose whether the intended functional chains remain complete and correct. Their lightweight feedback also helps agents localize violations without repeatedly reconsidering the entire design. These results demonstrate that explicitly modeling functional chains and enforcing them through executable specifications effectively improves repository-level feature development.

The additional architecture design and specification execution introduce moderate overhead. Compared with RTADev, \ourmethod increases the average cost by only \$0.01--\$0.03 per task, while consistently improving feature completion performance across different task scales. The additional interactions mainly come from executing functional-chain specifications and leveraging their violation feedback for iterative refinement, which provides more precise guidance during implementation. To further examine robustness across different backbone models, we additionally compare agent scaffolds using GPT-5.4-mini on the Lite split (Table~\ref{tab:gpt54mini_agent}). \ourmethod maintains the best performance, achieving 54.7\% \%Passed and resolving 5 tasks under the same backbone, even surpassing Codex, a dedicated coding agent developed by OpenAI for GPT series LLMs. These results suggest that the effectiveness of \ourmethod is robust across different backbone models rather than relying on a specific LLM.

\subsection{RQ2: Ablation Study}

\begin{table}[t]
\centering
\small
\setlength{\tabcolsep}{4.5pt}
\renewcommand{\arraystretch}{1.12}
\begin{adjustbox}{max width=\columnwidth}
\begin{tabular}{lccc}
\toprule
\textbf{Variant}
& \textbf{\%Passed}
& \textbf{\#Resolved}
& \textbf{Avg. Cost (\$)} \\
\midrule

\multicolumn{4}{l}{\textit{Specification Ablations}} \\
\quad w/o All Specification
& \score{62.6}{6.0}
& 7
& 0.09 \\

\quad w/o Behavior Specification
& \score{64.0}{5.2}
& 6
& 0.13\\

\quad w/o Architecture Specification
& \score{66.6}{5.7}
& 7
& 0.16 \\
\midrule
\multicolumn{4}{l}{\textit{Evidence-Grounded Functional Chain Ablations}} \\
\quad w/o Evidence
& \score{64.8}{5.7}
& 6
& 0.16 \\

\midrule
\ourmethod
& \bestscore{70.7}{5.1} & \textbf{9}

& 0.18 \\
\bottomrule
\end{tabular}
\end{adjustbox}
\caption{Ablation study on the FeatureBench Lite split. 
All variants use DeepSeek-V4-Pro as the base model. 
}
\label{tab:ablation_lite}
\end{table}

Table~\ref{tab:ablation_lite} reports the contribution of executable specifications and evidence-guided functional-chain construction. Removing all specifications decreases \%Passed from 70.7\% to 62.6\% and reduces the number of resolved tasks from 9 to 7, confirming that persistent executable constraints are important for long-horizon feature implementation. The two specification types provide complementary benefits. Removing the behavior specification causes a substantial drop to 64.0\% Passed and 6 resolved tasks, indicating that architecture connectivity alone cannot ensure correct outputs, boundary handling, or state transitions. Removing the architecture specification reduces \%Passed to 66.6\% and resolves 7 tasks, showing that behavior checks alone may still permit incomplete or unintended realization paths. Their combination therefore improves both path integrity and runtime correctness.

Evidence-guided functional-chain is also critical. Without explicit evidence grounding, \%Passed falls from 70.7\% to 64.8\%, while the number of resolved tasks decreases from 9 to 6. This suggests that requirement semantics and repository evidence help identify more reliable functional units and relations, thereby reducing unsupported architectural reasoning. Although the complete method incurs the highest average cost at \$0.18 per task, the increase is modest relative to the consistent gains in both partial progress and complete task resolution.

\subsection{RQ3: Executable vs. Textual Specifications}
\setlength{\textfloatsep}{10pt plus 2pt minus 2pt}
\begin{figure}[t]
    \centering
    \begin{subfigure}{\linewidth}
        \centering
        \includegraphics[width=0.95\linewidth]
        {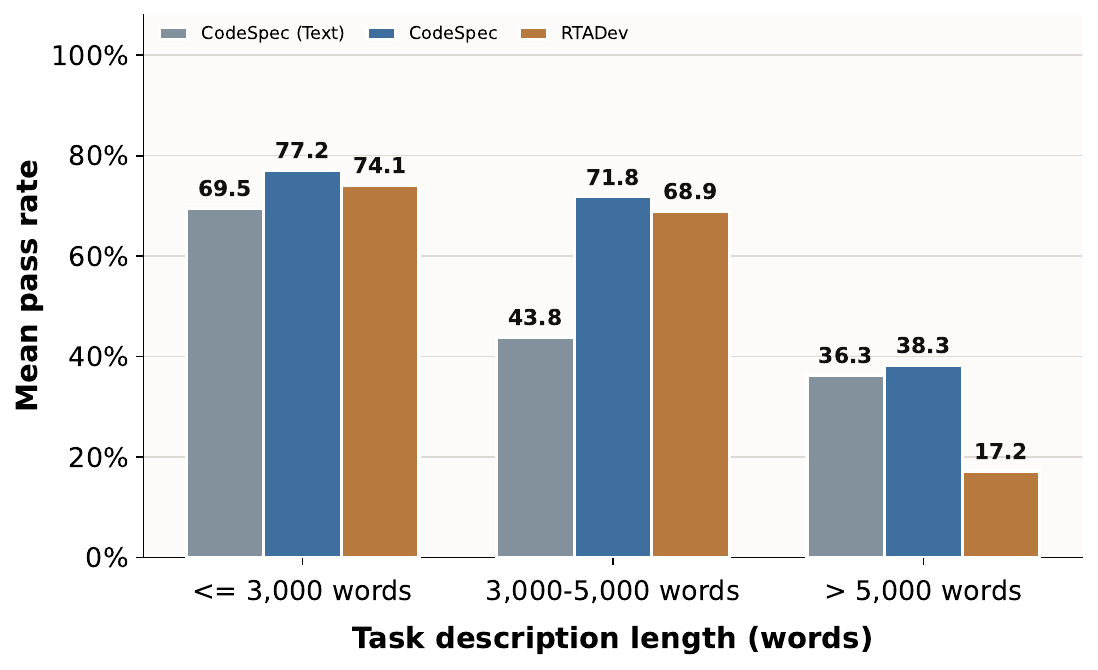}
        \caption{Pass-rate comparison across different instruction-length groups.}
        \label{fig:rq3_task_length}
    \end{subfigure}

    \begin{subfigure}{\linewidth}
        \centering
        \includegraphics[width=0.95\linewidth]
        {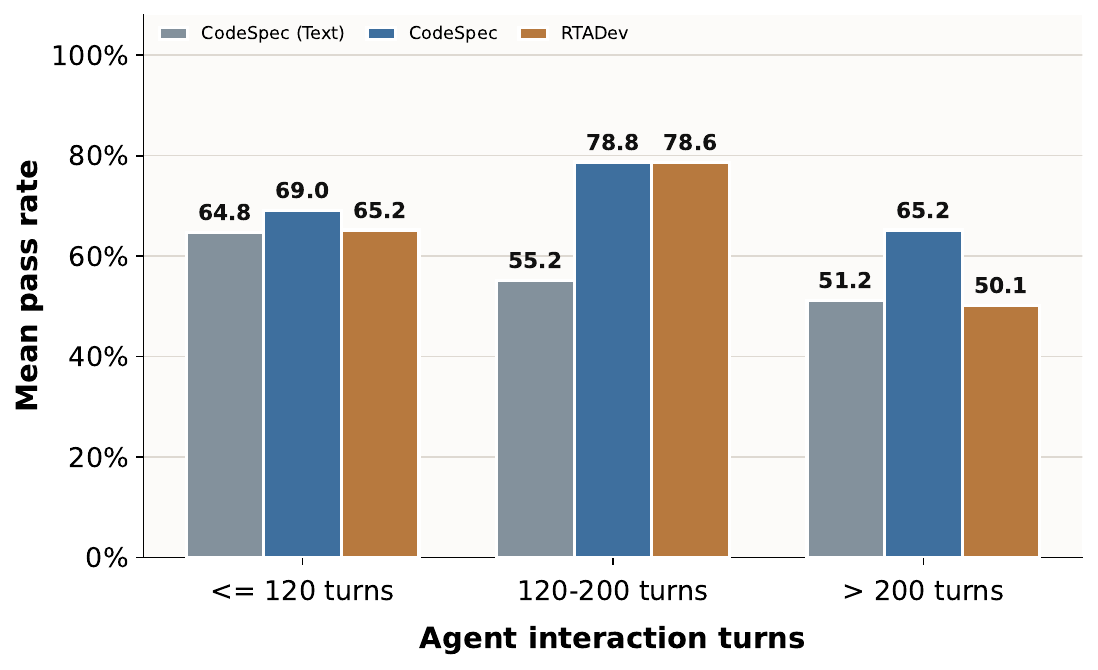}
        \caption{Pass-rate comparison across different agent interaction-length groups.}
        \label{fig:rq3_turns}
    \end{subfigure}
    \caption{Pass-rate comparison among textual specifications,
    executable specifications, and RTADev across tasks grouped
    by instruction length and agent interaction turns.}
    \label{fig:rq3_executability}

\end{figure}
We further investigate whether executable specifications preserve design constraints more effectively than textual specifications. \ourmethod (Text) retains the same evidence-grounded functional-chain construction as \ourmethod, but represents the resulting design as textual information rather than executable specifications. We compare it with \ourmethod and RTADev, another code agent based on textual design planning. We compared the three methods described above on the FeatureBench Lite split.

As shown in Figure~\ref{fig:rq3_executability}, the three methods perform relatively similarly on simpler tasks with instructions below 3,000 words or interactions under 120 turns. The gaps widen as task complexity increases: \ourmethod achieves 71.8\% on instructions of 3,000--5,000 words, compared with 43.8\% for \ourmethod (Text), and reaches 65.2\% beyond 200 turns, outperforming the two textual approaches by 14.0 and 15.1 points. Although textual plans also provide useful design information, placing all details in the context increases input size and the risk of overlooking requirements and design constraints. In contrast, \ourmethod provides lightweight executable feedback throughout interaction, offering more reliable guidance for requirement-intensive and long-horizon tasks.

\begin{table*}[t]
\centering
\small
\setlength{\tabcolsep}{5pt}
\renewcommand{\arraystretch}{1.15}
\begin{tabular}{lcccccc}
\toprule
\multirow{2}{*}{\textbf{Agent Scaffold}}
& \textbf{Easy}
& \textbf{Medium}
& \textbf{Hard}
& \multicolumn{3}{c}{\textbf{Overall}} \\
\cmidrule(lr){5-7}
& \textbf{\%Passed}
& \textbf{\%Passed}
& \textbf{\%Passed}
& \textbf{\%Passed}
& \textbf{\#Resolved}
& \textbf{Avg.Cost (\$)} \\
\midrule

OpenHands
& \score{48.6}{6.5}
& \score{26.2}{4.4}
& \score{9.4}{2.2}
& \score{26.4}{2.9}
& 2
& 0.10 \\

Claude Code
&  \score{62.0}{5.6}
&  \score{45.7}{4.6}
&  \score{19.3}{3.7}
&  \score{41.6}{3.2}
& 5
& 0.11 \\

Mini-SWE-Agent
& \score{58.0}{6.3}
& \score{35.0}{4.2}
& \score{18.7}{4.2}
& \score{35.5}{3.1}
& 4
& 0.08 \\

RTADev
& \score{61.3}{6.3}
& \score{47.4}{4.7}
& \score{16.3}{4.8}
& \score{41.3}{3.5}
& 7
& 0.14 \\

\ourmethod
& \bestscore{70.0}{5.6}
& \bestscore{51.7}{4.3}
& \bestscore{20.4}{4.7}
& \bestscore{46.6}{3.3}
& \textbf{8}
& 0.17 \\

\bottomrule
\end{tabular}
\caption{Comparison on NL2Repo using DeepSeek-V4-Pro (max thinking). Easy, Medium, and Hard correspond to repositories with $<$1.5K, 1.5K--4K, and $>$4K lines of code (LOC). Bold indicates the best result.
}
\label{tab:nl2repo}
\end{table*}
\subsection{RQ4: Greenfield Repository Development Generalization}
To assess generalization beyond repository evolution, we evaluate \ourmethod on NL2Repo-Bench~\cite{nl2repobench}, which contains 104 greenfield tasks that require generating a Python repository from a requirements document and an empty workspace. \ourmethod first creates a repository skeleton and functional units, after which the evolving codebase provides repository evidence for functional-chain construction and specification compilation. 

As shown in Table~\ref{tab:nl2repo}, \ourmethod achieves 70.0\%, 51.7\%, and 20.4\% \%Passed on the Easy, Medium, and Hard splits, outperforming the strongest baselines by 8.0, 4.3, and 1.1 percentage points, respectively. It also obtains the best overall \%Passed (46.6\%) and \#Resolved (8), demonstrating its generalizability to greenfield development. The larger gains on Easy and Medium tasks suggest that \ourmethod effectively organizes requirements into coherent implementations for repositories within approximately 4K lines of code, while the smaller gain on Hard tasks reflects the remaining difficulty of coordinating larger requirements for current agents. Although its additional specification and feedback steps slightly increase the average cost to 0.17, the performance gains indicate a favorable quality--cost trade-off.
\section{Discussion}
\subsection{Case Study}

\begin{figure}[h]
    \centering
    \includegraphics[width=1.05\linewidth]{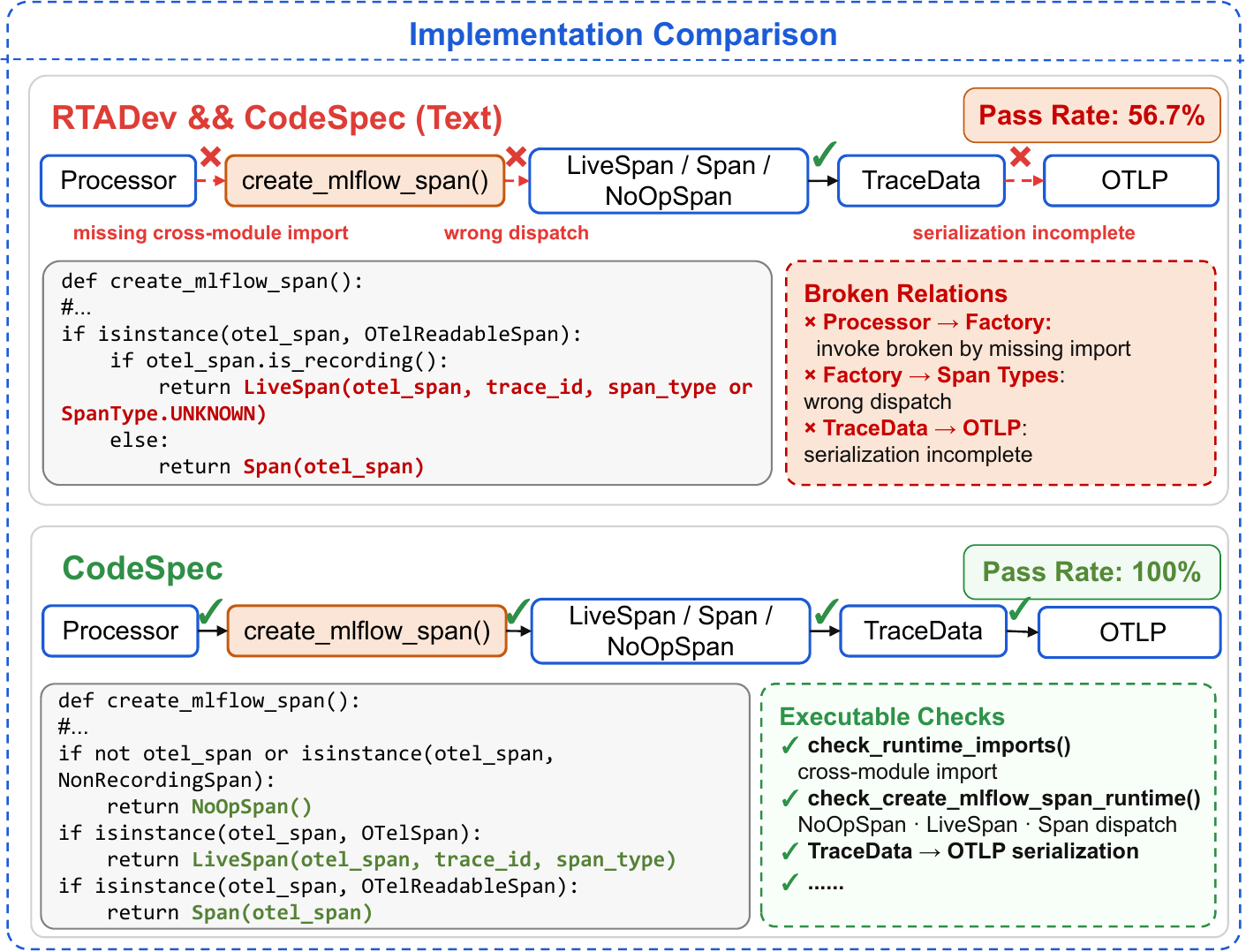}
    \caption{Case study of MLflow distributed tracing in Featurebench by using DeepSeek-V4-Pro.}
    \label{fig:case}
\end{figure}
Figure~\ref{fig:case} presents a representative MLflow feature-development task in Featurebench that requires the agent to preserve a complete span lifecycle across multiple modules. The RTADev and \ourmethod(Text) identify the major entities, but fail to preserve several cross-module relations during implementation. In particular, it dispatches spans based on \texttt{is\_recording()} rather than the required OTel span type, misses a required import between modules, and leaves the OTLP conversion path incomplete. Although many individual interfaces are implemented, these broken connections lead to an incomplete feature and a pass rate of 56.7\%.

In contrast, \ourmethod compiles the functional chain into complementary executable architecture and behavior specifications. Architecture feedback reveals whether the chain remains complete and connected, while behavior feedback identifies whether the connected path realizes the intended behavior correctly. This lightweight, targeted feedback allows the agent to localize and repair structural or behavioral violations during implementation. The case therefore shows that executable specifications go beyond textual design descriptions by continuously enforcing the functional chain and maintaining design--implementation consistency across long-horizon development.

\section{Conclusion}
We propose \ourmethod{}, a dual-specification approach for agentic long-horizon feature development. It constructs evidence-grounded functional chains from feature requirements and repository evidence, and compiles each chain into complementary executable architecture and behavior specifications. Architecture specifications preserve the integrity of feature realization paths, while behavior specifications verify runtime correctness along those paths. Together, they provide fine-grained and persistent feedback throughout implementation. Experiments on FeatureBench show that \ourmethod{} consistently outperforms representative coding agents, particularly on complex and long-horizon tasks. Moreover, results on NL2Repo-Bench further demonstrate its generalizability across repository-level development settings.

\bigskip

\bibliography{aaai2027}


\end{document}